\documentclass{egpubl}
\usepackage{eg2019}

\Education              
 \electronicVersion 

\ifpdf \usepackage[pdftex]{graphicx} \pdfcompresslevel=9
\else \usepackage[dvips]{graphicx} \fi

\PrintedOrElectronic

\usepackage{t1enc,dfadobe}

\usepackage{egweblnk}
\usepackage{cite}
\usepackage{tabu} 
\usepackage{booktabs}  
\usepackage{microtype}

\usepackage{ulem}
\makeatletter
\def\ifEmpty#1{\def\@temp{#1}\ifx\@temp\@empty}
\makeatother

\usepackage{xspace}

\usepackage{amsmath,amsfonts,amssymb}
\usepackage{url} 

\newcommand{\FG}[1]{Figure~\ref{#1}}

\newcommand{\TA}[1]{Table~\ref{#1}}

\newcommand{\shah}{{\textstyle \amalg{\kern-4.pt\amalg}}}

\newcommand{\tabitem}{~~\llap{\textbullet}~~}

\title[EG Education]%
      {Integrating Visualization Literacy into Computer Graphics Education Using the
Example of Dear Data}

\author[A. Krekhov, M. Michalski, and Jens Kr\"uger]
{\parbox{\textwidth}{\centering A. Krekhov$^{1}$, M. Michalski$^{1}$, and J. Kr\"uger$^{1}$
        }
        \\
{\parbox{\textwidth}{\centering $^1$Center for Visual Data Analysis and Computer Graphics (COVIDAG), University of Duisburg-Essen, Germany
       }
}
}

\begin{document}


\teaser{
  \centering
  \includegraphics[width=\linewidth]{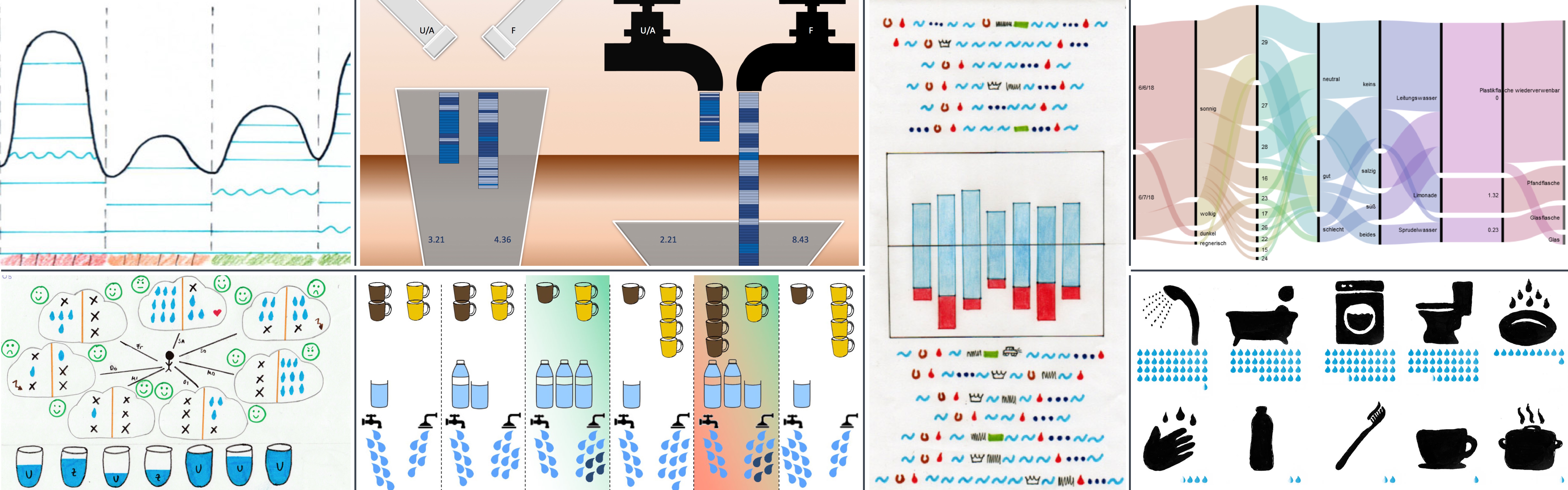}
  \caption{Results on the topic ``water'' showcase participants' divergent thinking approaches: day-by-day visualizations vs. aggregations, digital vs. analog, amount of tracked attributes, visual clutter, and differing topic interpretations.}
  \label{fig:teaser}
}

\maketitle
\begin{abstract}
The amount of visual communication we are facing is rapidly increasing, and skills to process, understand, and generate visual representations are in high demand. Especially students focusing on computer graphics and visualization can benefit from a more diverse education on visual literacy, as they often have to work on graphical representations for broad masses after their graduation. Our proposed teaching approach incorporates basic design thinking principles into traditional visualization and graphics education. Our course was inspired by the book \textit{Dear Data} that was the subject of a lively discussion at the closing capstone of IEEE VIS 2017. The paper outlines our 12-week teaching experiment and summarizes the results extracted from accompanying questionnaires and interviews. In particular, we provide insights into the creation process and pain points of visualization novices, discuss the observed interplay between visualization tasks and design thinking, and finally draw design implications for visual literacy education in general.


\end{abstract}  

\section{Introduction}

In our society, the demand for people being capable of creating meaningful and engaging visualizations rapidly outgrows the offer. The ability to visually transport a message to the broad masses requires a versatile education including computer graphics and visualization. However, we claim that our courses are often designed from a more academic point of view (which is not a critique!), and we tend to overlook the rather alarming, low level of visualization literacy of general audiences.

As researchers and lecturers, we sought for a possibility to enhance our courses in a way that would allow our students to create visualizations that are easy to understand, engage the viewers, and remain memorable over time. This paper presents our pilot lecture for undergraduate computer science students and reports results from our evaluations during its execution.

Our course design was particularly motivated by the book \textit{Dear Data}~\cite{lupi2016dear}, written by the two expert designers G. Lupi and S. Posavec, and its vivid discussion during the closing capstone at IEEE VIS 2017. The visualizations in that book were created by composing visualization knowledge and creativity. As a matter of course, such engaging results can be beneficial for a number of application areas, both entertaining and scientific in nature (cf. \FG{fig:teaser}). 

We wanted to know whether it is possible to enhance current visualization teaching such that students would be able to produce similar results to \textit{Dear Data} in terms of comprehension and engagement. In particular, our goal was to prevent students from any kind of tunnel vision that we often attribute to visualization novices. Instead, participants should rely on design thinking and hands-on exploration of the visualization space without being forced to proceed in the linear fashion that is often enforced by established tools.

Our teaching approach targets novices and practitioners with no background in design. The only methodology we borrow from that area is design thinking, or, more precisely, divergent thinking and brainstorming. Our hypothesis is that introducing design thinking is a valid way to increase the practitioners' awareness that there are far more ways than only line and bar charts to transport information. Design thinking motivates practitioners to alter their approach in problem solving, concentrating more on the solution to the problem instead of ``over analyzing'' the problem. The key feature of design thinking is to develop several prototypes that are used to find the final best fitting solution. It is more a creative way of solving a problem, encouraging practitioners to actually ``do'' something. Furthermore, such solution-oriented thinking should prevent the visual representation that is imposed by the software currently in use.

Designing such a course is an iterative approach and heavily relies on knowledge about the status quo, i.e., how novices actually proceed when they face a visualization task. Hence, we launched a teaching experiment to investigate the visualization creation pipeline and, more importantly, its interplay with the design thinking methodology. As a starting point for our teaching experiment, we chose the \textit{Dear Data} approach, i.e., our participants had to track certain data each week and create a meaningful visualization at the end of the week. The tracked data could be, for example, \textit{``water consumption''} or \textit{``music''}. Hence, participants had a large impact on the dataset generation and usually had a deep understanding of the data because it originated from their daily lives.

Our 12-week pilot course was accompanied by biweekly one-on-one semistructured interviews and monthly anonymized online questionnaires. During the 12 weeks, i.e., one semester, we varied the underlying homework conditions by letting the participants work alone and in groups of varying size to see how the conditions influence the visualization creation process. 

To summarize, this paper contributes  a preliminary course design, including the discussion of its limitations and drawbacks, and lessons learned along the way. We provide insights into the thinking process and the visualization pipeline of novices and expect that other visualization teachers can substantially benefit from these first steps into the systematic strengthening of visual literacy in education.

 \begin{figure*}[t!]
 \centering 
 \includegraphics[width=2.0\columnwidth]{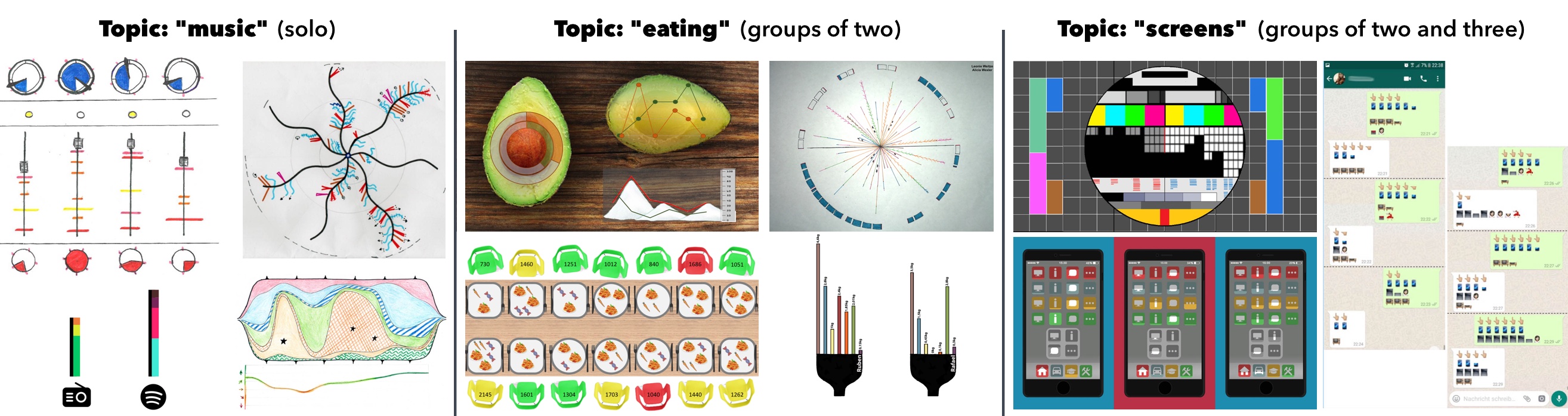}
 \caption{An extract from the produced results ordered by topic and group size. As can be seen here (and also confirmed by the questionnaire), visual metaphors played an important role regarding the final visualization. Also note that working in groups resulted mostly in digital submissions, as presented in \FG{fig:tools}.}~\label{fig:comp}
\end{figure*}



\section{Related Work}

%
%

Our experiment is based on a combination of traditional visualization approaches and certain elements of design thinking. In the following subsections, we briefly discuss visualization creation models, core aspects of infographics, and related studies with visualization novices, and we provide a brief introduction into design thinking for nondesigners.

\subsection{Visualization}


The research on understanding visualization creation has resulted in a plethora of---partially contradicting---taxonomies, assumptions, and related discussions in the visualization community~\cite{bertin1983semiology,Card:1999:RIV:300679,Chi:2000:TVT:857190.857691,1382903}. In this regard, we mention the work by Shneiderman~\cite{shneiderman2003eyes} on the systematic visualization classification and the related prominent mantra: ``overview first, zoom and filter, then details-on-demand'', which is often a helpful thought process during visualization creation. Another reasonable starting point for visualization apprentices is the so-called periodic table by Lengler et al.~\cite{Lengler:2007:TPT:1712936.1712954}: the authors summarized over 100 visualization methods in order to structure available mechanisms and to create a useful asset toolbox for researchers and practitioners. 

Instead of such data-centered approaches, the high-level taxonomy by Tory et al.~\cite{1382903} focused rather on human factors in visualization processes. By inspecting and grouping the algorithms chosen during the visualization process, the authors illustrated the interplay between design and user models. Another different classification approach based on vision perception was proposed by Rodrigues et al.~\cite{rodrigues2006reviewing}. The authors utilized the knowledge about preattentive stimuli to describe visualization elements. In this context, we point readers to the work by Healey et al.~\cite{Healey:2012:AVM:2225054.2225226} for a better understanding of the underlying preattentive features such as motion or binocular rivalry~\cite{krekhov2019deadeye} and their role in visualization.

To design a successful visual representation, we also need to know how such graphics are perceived. Harrison et al.~\cite{Harrison:2015:IAD:2702123.2702545} examined the overall appeal of infographics and measured how quickly the aesthetic impression is formed. Hereby, three main aspects are often emphasized: comprehension, retention, and appeal (e.g., Lankow et al.~\cite{lankow2012infographics}). Appealing visualizations often require artistic methods, and there is an ongoing debate about how much of art is actually useful. For instance, Tufte~\cite{Tufte:1986:VDQ:33404} criticized chart junk, whereas Bateman et al.~\cite{Bateman:2010:UJE:1753326.1753716} showed that visual embellishments have a positive effect on retention. As a follow-up to that debate, Borkin et al.~\cite{6634103} executed a large-scale experiment with more than 2000 visualizations to study how exactly we can make our visualizations more memorable. 

Related to our evaluations are the following experiments: Bigelow et al.~\cite{Bigelow:2014:RDD:2598153.2598175} extensively studied professional designers and their workflow regarding visualization. The authors summarized current pain points and proposed patterns for the creation of tailored tools for these experts. In contrast, Grammel et al.~\cite{5613431} focused on novices. The authors pointed out the most prominent activities during visualization specification and also identified main barriers that occur during the process. Our work picks up on that idea, but shifts the focus towards aesthetically appealing and engaging data-driven infographics. 

To convey initial visualization knowledge to a broader audience, we recommend the article by Heer et al.~\cite{Heer2010ATT}. The authors conduct what they call a ``a brief tour through the visualization zoo'' to expose common visualization techniques. Similar to our teaching approach, the authors motivate, readers to explore the visualization design space to strengthen solution-oriented visualization creation.

Regarding ongoing research in visualization education, we emphasize the work by Roberts et al.~\cite{roberts2016sketching} on the ``Five Design-Sheet Methodology'' that establishes a more formalized design process for lo-fidelity prototypes. Similar to our goals, the authors underpin the necessity to explore different possibilities and to evaluate their effectiveness before narrowing down to a single visual representation. Also aligned with our course direction is the so-called ``VizItCards''~\cite{he2017v} toolkit, which should be seen as a workshop designed for graduate infovis classes. The authors focused on five learning objectives: design, ideate and compare, collaborate, apply, and synthesize. Our evaluation results show a similar execution process, but a less linear way of progression, e.g., rejecting the whole concept in the apply stage and returning to design.

\subsection{Design Thinking}

Classically, design thinking can be described as a solution-focused problem-solving approach~\cite{Pusca:2018:DTAPS, DMJ:DMJ12001} that was well studied in the design domain~\cite{Goldschmidt:1994:VDTKA,Cross:1996:ADA}, but is also highly supported outside the design context within fields like IT~\cite{Brooks:2010:DDEFCS}, allowing users to view and solve their problems in a new way. Contrary to what is commonly believed, analytic aspects are also considered in design thinking as stated by Razzouk and Shute~\cite{Razzouk:2012:WDTI}: 

\begin{quote} 
Design thinking is generally defined as an analytic and creative process that engages a person in opportunities to experiment, create and prototype models, gather feedback, and redesign.
\end{quote}

Inspiration, ideation, and implementation are the iterative stages identified within the available design thinking paradigms. Although design thinking is described differently~\cite{Razzouk:2012:WDTI,Buchanan:1992:WPDT} and, as pointed out by Dorst~\cite{Dorst:2011:CDTA}, could be seen as not well conceptualized, these three stages are often used as a basis~\cite{Palacin-Silva:2017:IDTISECC}. Inspiration focuses on the empathy of the client to support the search for solutions. During the ideation stage, various solutions are developed and tested. A selection of these solution sets is then utilized during the implementation stage to create prototypes, which are then refined iteratively until the best solution is found. As a summary, Lindberg et al.~\cite{Lindberg2011} and Pusca et al.~\cite{Pusca:2018:DTAPS} describe design thinking as two phases: exploring the problem space (inspiration) and inspecting the solution space (ideation and implementation). Similarly, our course encourages participants to explore their gathered data and to try out different visual representations to create a meaningful and engaging end result.

\section{Developing the Course Curriculum}

Our preliminary course design mainly targets university students in computer science or related areas. We assume participants have no or little knowledge in visualization and no design background. Although we introduce certain basic elements of design thinking, the design portion is kept minimal, and, therefore, the course execution is reasonable for the majority of cg/vis instructors. Our teaching approach is not meant to replace current courses and should be considered as a potential extension. We suggest to administer the course before the main cg/vis lectures to offer a first hands-on experience in visual literacy. The primary course objective is to \textbf{learn how to turn manifold datasets into  meaningful and engaging visualizations for a broader audience}. We split the objective into the following components:

\begin{itemize}
  \setlength{\itemsep}{1pt}
  \setlength{\parskip}{0pt}
  \setlength{\parsep}{0pt}
\item \textbf{understanding data:} learn how to collect, explore, group, classify, and dismiss data
\item \textbf{visualizing data:} learn how to map data to an appropriate visual representation (i.e., the basics from traditional visualization teaching courses)
\item \textbf{design thinking:} learn to combine divergent thinking and visualization techniques to transport key messages about the data
\end{itemize}

\begin{table}[tb]
  \caption{Assignment topics and according group sizes during our pilot course. The students could freely interpret the topic and decide what kind of associated data to track and to visualize.}
  \label{tab:topics}
  \scriptsize%
	\centering%
  \begin{tabu}{%
	r%
	*{7}{c}%
	*{2}{r}%
	}
  \toprule
   Topic & Assignment Duration & Group Size   \\
  \midrule
	New Acquaintances & 1 week & alone   \\
  Seating & 1 week & alone   \\
  	Music & 1 week & alone   \\
  Colors & 1 week & alone   \\
  	Water & 2 weeks & alone   \\
  Eating & 1 week & 2 students   \\
  	Screens & 1 week & 2-3 students   \\
  Joy & 2 weeks & 3-4 students   \\
  	Infographics Contest~\cite{contest} & 2 weeks & free choice   \\
  \bottomrule
  \end{tabu}%
\end{table}

Especially the last subgoal is a key difference to common visualization courses. From our prior teaching experience, we noticed that many students struggle to find an appropriate kind of visualization because of their problem-solving pipeline. More precisely, students tend to pose questions such as ``How can I fit the data into a line chart/bar chart'' instead of asking ``What representation is suited best to transport certain information''. Students usually try to find the best solution by relying only on their gathered data instead of producing several  representations and using them as a basis to decide which fits best the current task. In other words, we observed that less experienced practitioners often perform an overhasty selection from a small number of basic visualization techniques. In our opinion, such a quick convergence often results in the sort of tunnel vision that hinders practitioners from revising their primary decision and from reconsidering alternative visual representations.

We wanted find out whether and how divergent thinking can be integrated into the education pipeline to solve the outlined issue. In our pilot course, we integrated the three stages of design thinking (inspiration, ideation, and implementation) as proposed by Palacin-Silva et al.~\cite{Palacin-Silva:2017:IDTISECC}. As a starting point, we applied the \textit{Dear Data} procedure: students should gather data from their daily lives on a weekly basis and create suitable visualizations. \textit{Dear Data} itself is an extensive example of creative and diverse visualization techniques. Hence, we assume that the underlying procedure---and not only the skills of the authors---might have a positive impact on divergent thinking. Furthermore, we argue that collecting and visualizing personal data during education has certain benefits over working with predefined datasets. Students tend to understand their data, and, according to our observations, experience significant motivation to explore and visually present their findings.
After collecting their data, they were directly shifted to the inspiration stage to view their data and search for some opportunities and interesting features that could answer the question: ``What is the message of my data?'' In the ideation stage, students should make their ideas tangible, selecting different types of representation for each idea to develop a basis for the final stage. In this stage, students iteratively evaluate their solutions to decide on one idea and on one visual representation of that idea. The benefits of this pipeline are the possibility to visually review several ideas and representations simultaneously. We assume that this solution-focused problem-solving approach helps novices with less experience to derive meaningful visualizations.

The following course routine remains roughly the same during the course duration. For each session, students have the assignment to create an engaging, comprehensive visualization based on the data that they tracked for a given topic as shown in \TA{tab:topics}. Participants can interpret the topic the way they want and freely choose the subdimensions to track. For instance,  ``music'' could include genres, current mood, current place, or the hardware the music is played on. This freedom explicitly enforces the students to carefully think about the data and possible hypotheses and correlations they want to discover. Optimally, students make these data-related decisions based on brainstorming, which gently pushes them into a divergent thinking mode. In other words, the students' main goal is to extract core messages from data and communicate them in a visual way.

 \begin{figure}[t]
 \centering 
 \includegraphics[width=\columnwidth]{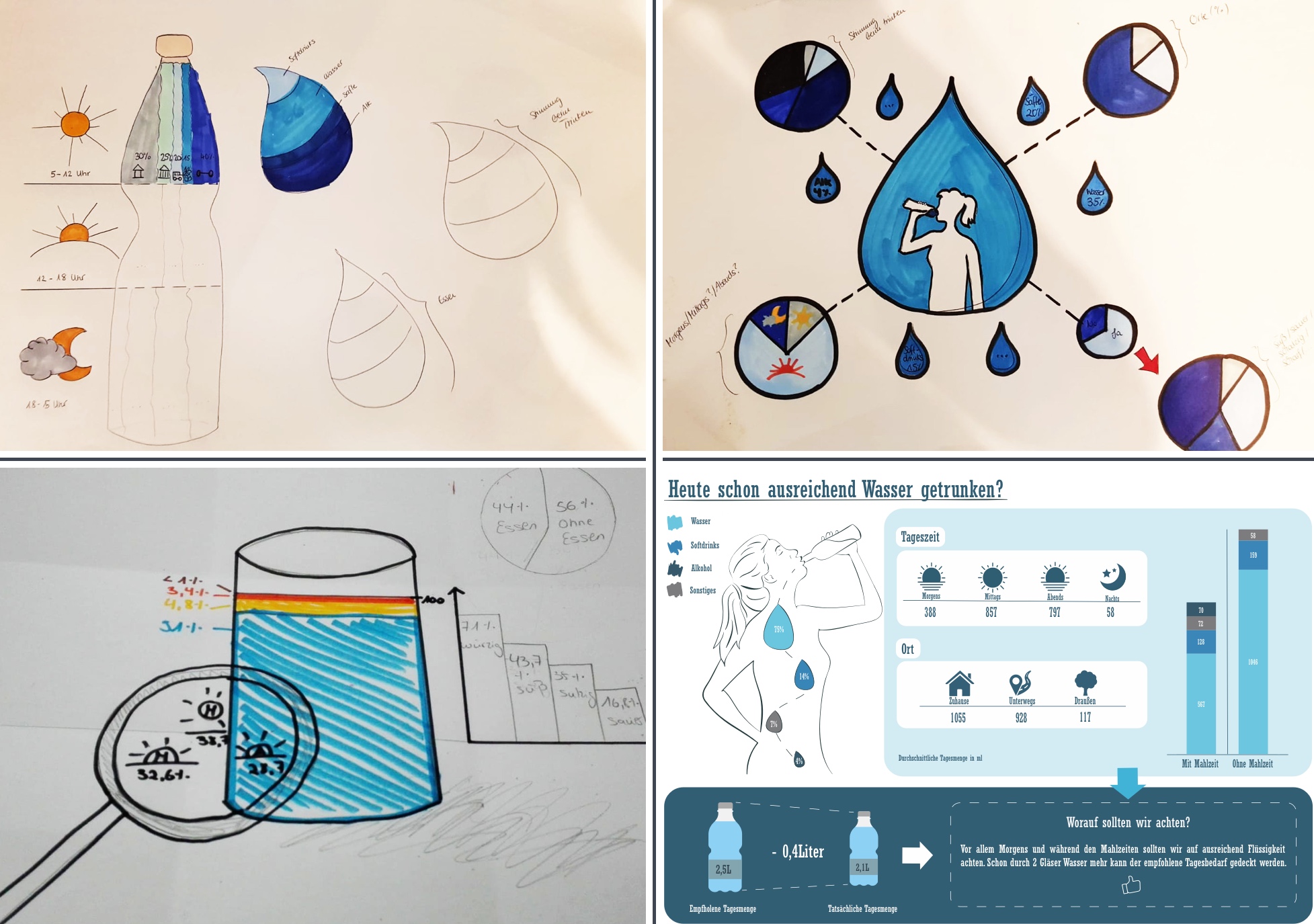}
 \caption{Students should submit their intermediate prototypes to demonstrate how their ideas evolved over time. Groups often started by producing 3-6 initial lo-fidelity concepts and then merging selected components into intermediate results.}~\label{fig:proto}
\end{figure}

\begin{table}[tb]
  \caption{Structure of our online questionnaire. Most of the 103 questions were based on qualitative insights from the semistructured interviews during the course.}
  \label{tab:questions}
  \scriptsize%
	\centering%
  \begin{tabu}{p{1.8cm} p{6.2cm}}

  \toprule
   Category & Example Questions (shortened)    \\
  \midrule
  \addlinespace
  \textbf{Visualization Creation Pipeline} & 
  24 Questions targeting the whole creation process and particular steps, e.g.: \newline
  \textit{
  \tabitem describe stepwise your weekly approach\newline
  \tabitem describe how your process evolved during the course \newline
  \tabitem which step did you enjoy most and why \newline
  \tabitem which step was the biggest challenge for you and why \newline
  \tabitem which was the most time-consuming step and why}
\\
  \addlinespace

  \textbf{Data Tracking} & 
    8 Questions related to data acquisition, e.g.: \newline
  \textit{
    \tabitem how did you acquire data (digital/analog) \newline
  \tabitem which tools did you use for data acquisition \newline
  \tabitem how frequently did you acquire data \newline
   \tabitem did you track enough dimensions
   }
\\
  \addlinespace
  
    \textbf{Data Analysis} & 
    9 Questions about data exploration, e.g.: \newline
  \textit{
     \tabitem how much time did you spend on analyzing your data\newline
  \tabitem which tools did you use to analyze the data\newline
  \tabitem have you ever adjusted/modified your data to better fit the visualization}
\\
  \addlinespace

    \textbf{Core Message} & 
    8 Questions about the message to be transported, e.g.: \newline
  \textit{
  \tabitem at which point did you establish your message(s) \newline
  \tabitem how difficult was it for you to establish a message \newline
  \tabitem how often and why did you re-adjust your message}
\\
  \addlinespace

      \textbf{Visualization} & 
    25 Questions related to the vis. creation, e.g.: \newline
  \textit{
  \tabitem how many intermediate drafts have you created on average \newline
  \tabitem which tools did you use to create the drafts \newline
  \tabitem describe your common iterations during the drafting stage\newline
  \tabitem which tools did you use for your final vis. \newline
  \tabitem how did you decide which colors to utilize \newline
  \tabitem what aspects of the final vis. were the most important for you \newline
  \tabitem what were the common reasons to re-do your vis.
  }
\\
  \addlinespace

        \textbf{Teamwork} & 
    29 Questions related to the second half of the course when students had to work in group, e.g.: \newline
  \textit{
  \tabitem describe your creation process in team \newline
  \tabitem describe whether and how teamwork positively influenced your results \newline
  \tabitem rate the data tracking/acquisition/analysis/ complexity compared to your previous work \newline
  \tabitem describe whether and how your process was different when working with friends vs. working with ``strangers'' \newline
  \tabitem in future, would you rather prefer visualizing alone or in group and why
  }
\\
  \addlinespace

  \bottomrule
  \end{tabu}%
\end{table}

 \begin{figure*}[t!]
 \centering 
 \includegraphics[width=1.9\columnwidth]{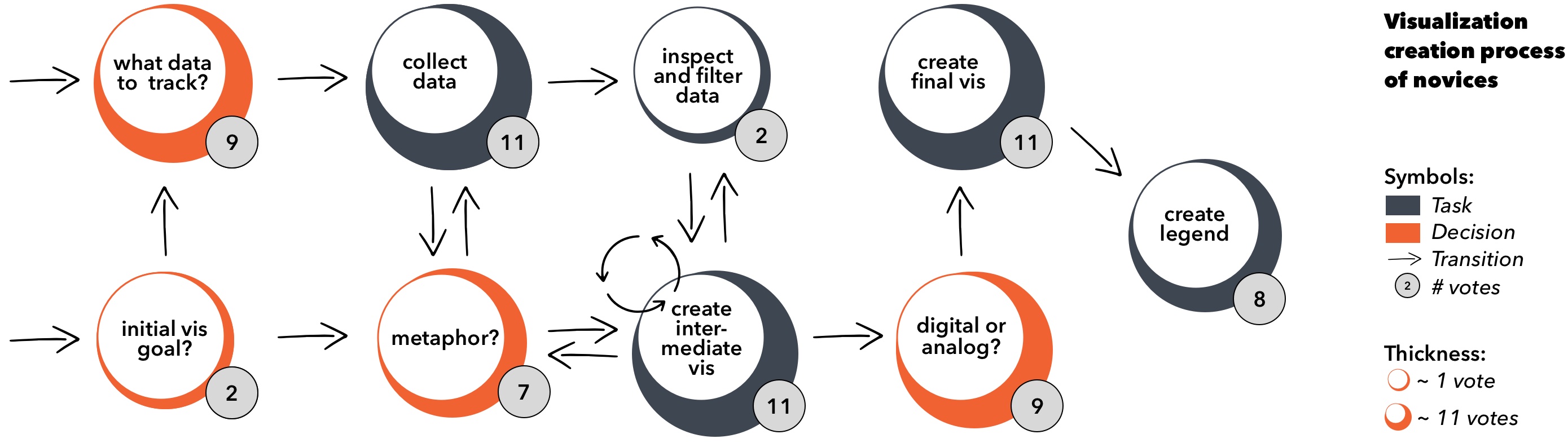}
 \caption{The nonlinear, iterative creation approach based on aggregated reports of our subjects.}~\label{fig:pipe}
\end{figure*}

As we want to keep our influence on that solution-oriented process as minimal as possible, we explicitly do not impose any limitations on how the data should be tracked or how the final visualization should be created. The only requirement that we communicate is the submission format for the visualization result: DIN A4, either digital or analog, and an additional page for the legend, if needed. In addition to the final result, students should also submit their early and intermediate prototypes as shown in \FG{fig:proto}. That additional material helps to understand the decision making and the divergent thought process of the students and serves as a discussion outline during final presentations.

After the submission, students have to present their outcomes during our weekly meeting. From the students' point of view, the weekly meeting slot has two main purposes. First, they have to present their work and get the opportunity to learn from other outcomes. Second, the students receive feedback from the lecturers. Our feedback focuses on ad hoc explanation and background on the utilized visualization techniques. In particular, we talk about the best practices, benefits, and drawbacks of a certain method, e.g., of a radar chart. Note that we do not postulate that our a posteriori teaching approach is superior to a priori methods, i.e., outlining the visualization methods before the hands-on part. Our main motivation is rather to explore where unbiased novices struggle and to see how diverse their results could potentially be before an extensive theoretical briefing. For future courses and pilot experiments, we suggest to enhance the course design by a small amount of frontal instructions and explore how that influences the creation pipeline.


%
%
%
%

\section{Pilot Experiment}

In the following section, we summarize the realization of our concept as a 12-week teaching experiment and our applied research apparatus. We were driven by four particular questions: 

\begin{itemize}
  \setlength{\itemsep}{+--+-++1pt}
  \setlength{\parskip}{0pt}
  \setlength{\parsep}{0pt}
\item How do novices proceed, i.e., what is their pipeline?
\item How do design thinking and visualization creation intertwine?
\item How does collaboration influence the process?
\end{itemize}

Our experiment was led by one visualization faculty member and two senior researchers (with background in cognitive science and computer science), all with prior experience in visualization education (Computer Graphics, and InfoVis/SciVis courses). We had 11 participants (8 females, 4 males), aged 20 to 31 ($M = 22.73,\,SD = 3.95$). At the beginning of the course, all participants reported that they were beginners regarding visualization and had no background in design education. The subjects studied either computer science (5 subjects) or applied cognitive and media science (6 subjects).

During the first 6 weeks, subjects worked alone, whereas in the second half, we formed groups of two to four participants and asked each group to create one collective visualization. We did not impose any restrictions about whether the tracked data should be aggregated for the whole group or remain visible for each individual. We also offered the students the opportunity to participate in an infographics contest~\cite{contest} and to apply their gathered skills on external datasets. A more detailed overview of our topics can be found in \TA{tab:topics}.

We surveyed the participants in two rounds of one-on-one semistructured interviews during the course, i.e., after 4 and after 8 weeks, and one anonymized online questionnaire at the end of the experiment. The questionnaire structure is presented in \TA{tab:questions}. The subsections, particular questions, and research directions are mostly based on the observations of the preceding semistructured interviews.

\section{Results And Discussion}
\subsection{Data Exploration Issues}

%

%
%

%

In order to extract similarities in the creation processes, we asked all participants to describe their weekly approaches in a step-by-step fashion. The resulting visualization pipeline is depicted in \FG{fig:pipe}. One important observation is that participants often ignored the data analysis step. Accordingly, subjects rarely filtered or sorted out data. This finding is also confirmed by the given answers in the data analysis subsection of the questionnaire. We suppose that skipping the data analysis step is the most relevant issue and should be addressed more explicitly in our next course iteration. 

Although the participants created several intermediate visualizations, the reason for iteration was purely design-based, i.e., these visualizations were never used as a tool to examine the data and search for correlations. Instead, participants reported the reason was to \textit{``unveil interesting trends by intuition, by looking at raw data or data tables, and by manually computing mean values'' (P3; 9 answers)}. Similarly, 10 subjects stated that they had formulated the main message or goal of their visualization during or even before the data acquisition. In other words, we should explicitly advise novices to thoroughly explore the underlying data. In most cases, standard tools such as Excel would minimize errors during acquisition and also provide useful default visualizations that help users to discover interesting trends and correlations.

 \begin{figure}[t]
 \centering 
 \includegraphics[width=0.9\columnwidth]{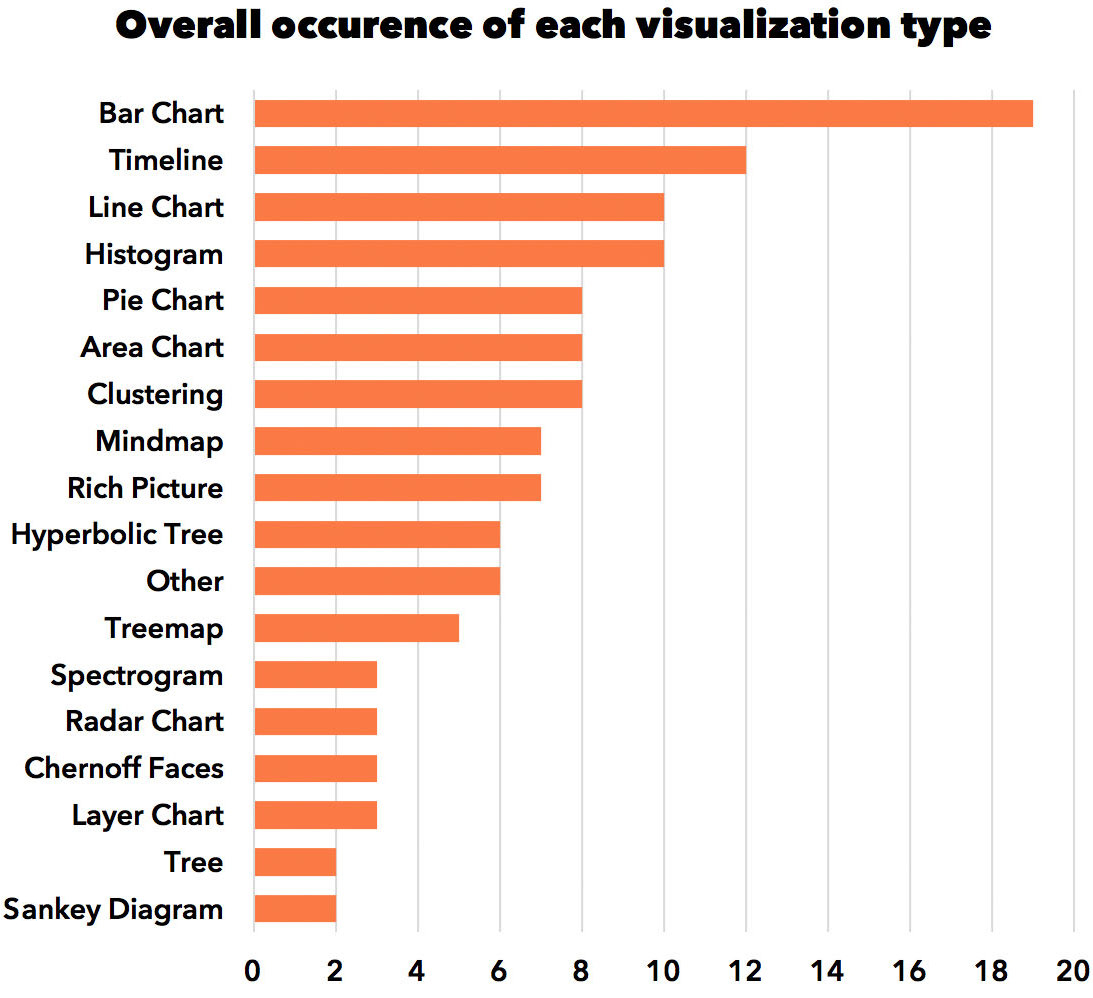}
 \caption{Students utilized a plethora of visualization techniques throughout the pilot experiment, i.e., these techniques tend to be familiar and comprehensible even for beginners.}~\label{fig:count}
\end{figure}

\subsection{Impact from Design Thinking}

All participants reported that creating the final visualization was most challenging and most time-consuming, but also most exciting. Surprisingly, they do not want any automated tools for this step, and mostly \textit{``enjoy the freedom of choice between hand drawings and digital creations'' (P10; 8 answers)}. Overall, students reported performing a number of quick iterations on different visualization approaches before going in-depth regarding the final creation. Hence, the introduction of design thinking elements indeed motivates students to explore the visualization design space and to experiment with a diversity of techniques and metaphors (cf. \FG{fig:comp}). 
 


We also counted and classified the resulting visualizations by projecting the outcomes onto canonical visualization techniques where applicable. During our classification process, we mostly relied on the periodic table by Lengler et al.~\cite{Lengler:2007:TPT:1712936.1712954}. The resulting distribution of visualization techniques is given in \FG{fig:count}. That summary underlines the diversity of utilized methods, even without a prior introduction by the instructors. In other words, these visualizations could be considered to be very intuitive and familiar to beginners.

\begin{figure}[t]
 \centering 
 \includegraphics[width=\columnwidth]{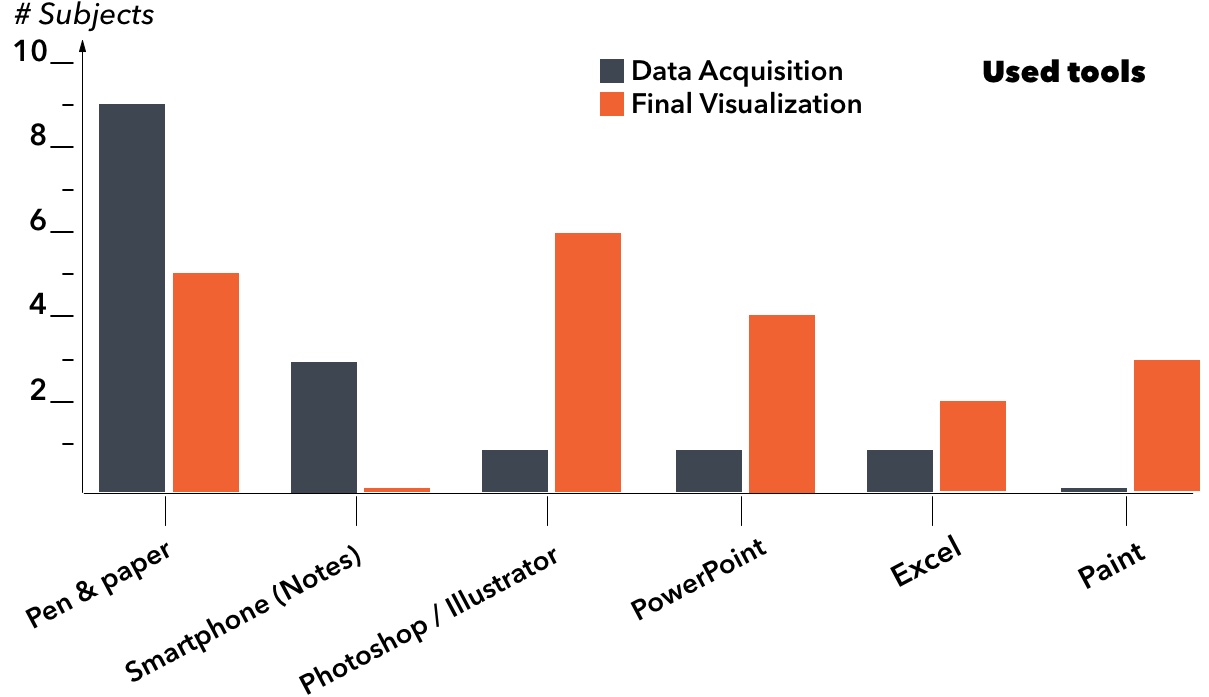}
 \caption{Tools used by subjects for data tracking and visualization.}
 \label{fig:tools}
  \vskip -3mm
\end{figure}

When asked about visualization-specific issues, multiple subjects reported a reverse learning effect: \textit{``I think I needed more time each week because I ran out of visualization ideas''}. Other participants mentioned limited craftsmanship: \textit{``It happens often that you mix up certain symbols or choose the wrong colors'' (P8; 5 answers)}.

We conclude that design thinking motivates novices to experiment with a broader range of visualization methods. That, in return, potentially increases the probability of finding a well-suited visual representation. We should not bind apprentices to a particular visualization tool as outlined by \FG{fig:tools}, and should provide enough example visualizations as a source of inspiration.

\subsection{Visualization Quality from Students' Perspective}

In retrospect, we asked all participants what they thought is important for a good visualization. The most prominent factors according to the participants are \textit{appeal}, \textit{metaphor}, and \textit{comprehension} (intended goal). Regarding aesthetics, the choice of colors and shapes played a big role: \textit{``I tried to choose highly differing colors for different attribute types, and varied the gradation to further subdivide an attribute'' (P4; 6 answers)}. Also, some participants preferred a \textit{``minimalistic and intuitive design'' (P11; 3 answers)}.

Regarding the comprehension aspect, subjects told us that their visualization \textit{``should reveal important trends, differences and similarities.''} All subjects indicated that \textit{``others should be able to quickly grasp the important points and understand the core message of the visualization'' (P5, 10 answers)}. However, no one mentioned the memorability aspect of the visualization. However, we observed that memorability, which is the third pillar of good infographics~\cite{lankow2012infographics} along with comprehension and appeal, is often neglected. We suggest to emphasize the importance of memorable representations~\cite{6634103,Bateman:2010:UJE:1753326.1753716}, as this aspect seems to be the least intuitive and present from the students' point of view.

\begin{figure}[t]
 \centering 
 \includegraphics[width=\columnwidth]{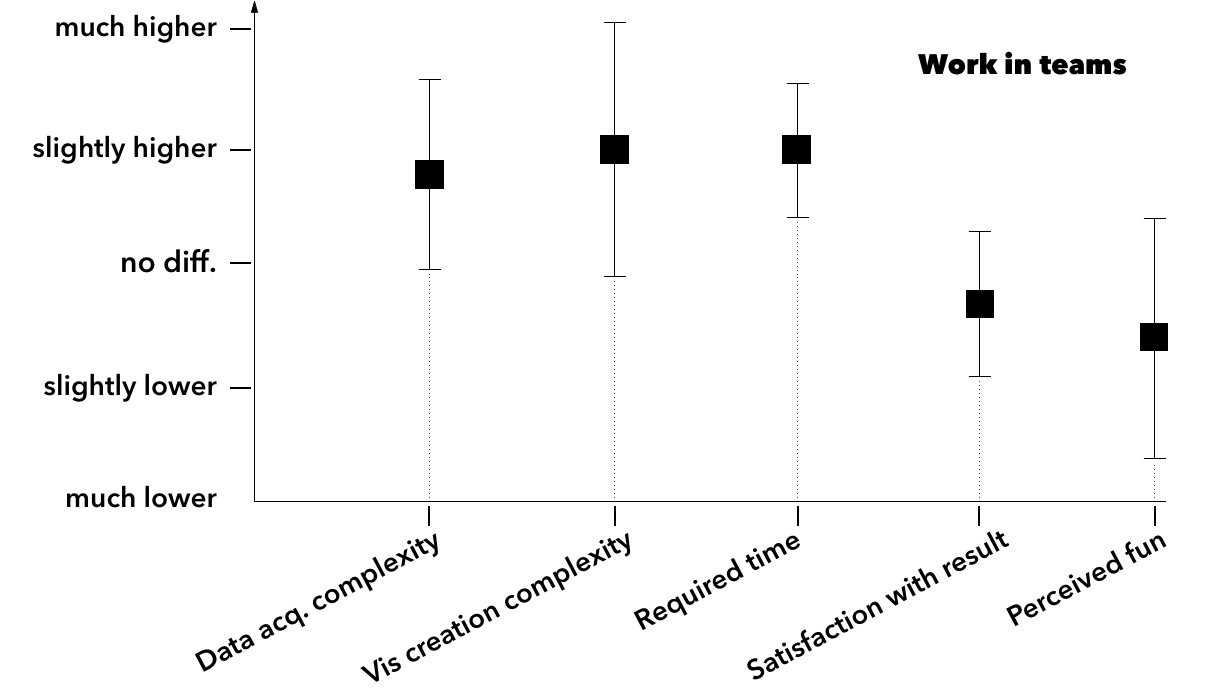}
 \caption{Perception of teamwork compared to working alone.}
 \label{fig:team}
\end{figure}

\subsection{Visualization as a Group Assignment}

Our participants had to create the last six visualizations in groups of varying size. The exact group sizes and according topics are shown in \TA{tab:topics}. Our final questionnaire regarding the teamwork part included a number of relational questions and, thus, mostly accounted for the experienced changes in the visualization creation process. 

As can be seen in \FG{fig:team}, the majority reported that the collaborative process is more complex, time-consuming, and less fun. One particular reason is that subjects had to make compromises: \textit{``Our group result was always sort of in-between without a clear structure, and I felt that creativity suffered a lot by making too many compromises'' (P10; 3 answers)}. The group size also influenced the subjects' perception on collaboration. In contrast to groups of three or four, participants stated that  \textit{``Working in a group of two was much more efficient thanks to the improved communication and more different ideas that could be evaluated'' (P8; 4 answers)}. 

We noted that nearly all team creations were done digitally. In contrast, about half of the solo results were created in an analogous manner, as shown in \FG{fig:tools}. Participants stated that the digital way allowed easier modifications of ongoing work: \textit{``If someone painted a green car and we disliked the color, he did not have to repaint it by hand'' (P9; 8 answers)}. The toolchain diversity was also mentioned: \textit{``Everyone was skilled at different tools and we often worked with multiple programs at the same time'' (P4; 7 answers)}.

\begin{figure}[t!]
 \centering 
 \includegraphics[width=\columnwidth]{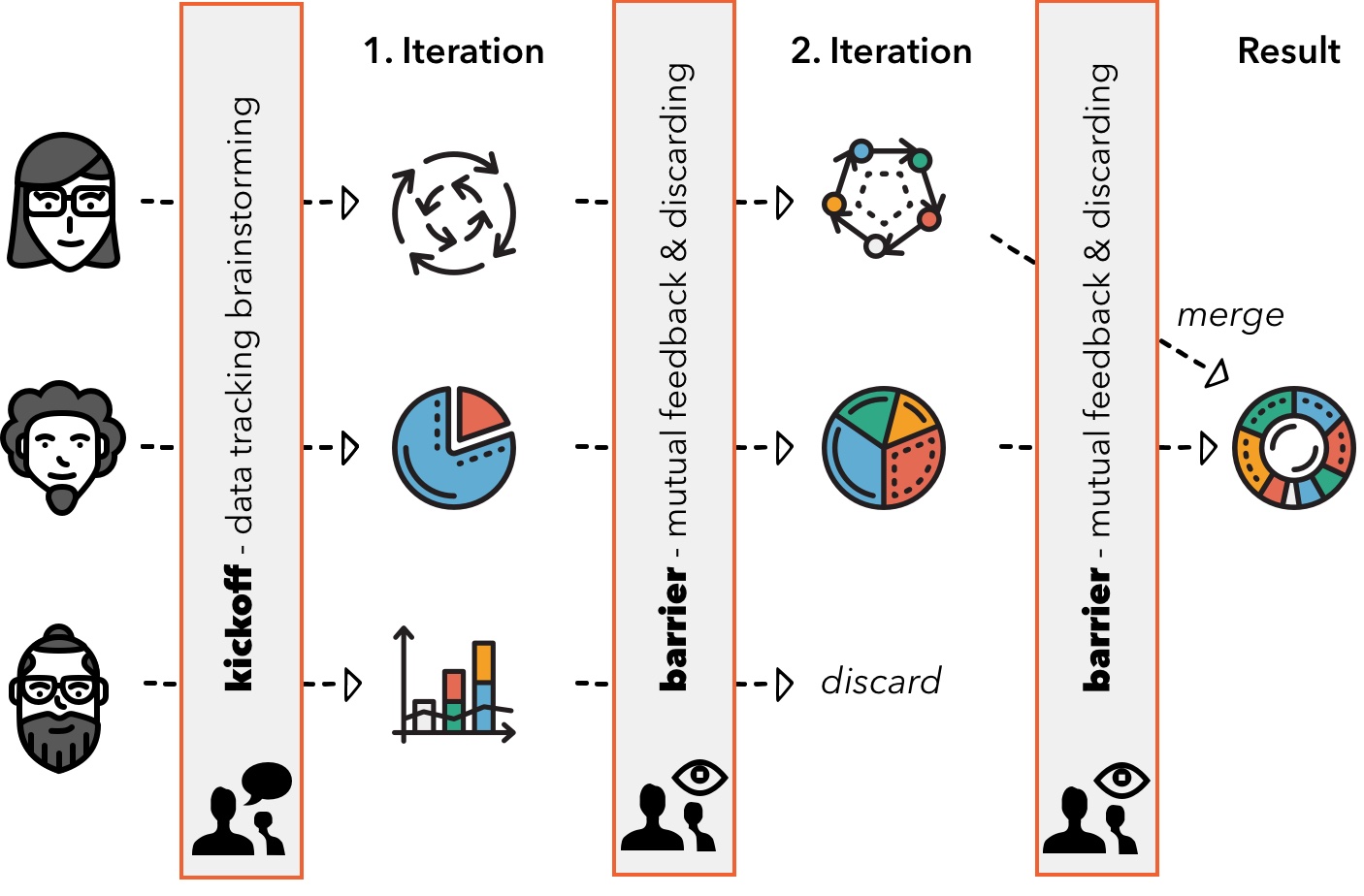}
 \caption{Groups often relied on a barrier-based approach. Instead of agreeing upon a single visualization, each member created several iterations of own ideas. During their meetings, members gave mutual feedback and decided whether a particular approach should be dropped or merged.}
 \label{fig:barrier}
\end{figure}



A particularly interesting observation was what we called a \textit{barrier-based approach}, as outlined in \FG{fig:barrier}. Groups repeatedly reported a quite uncommon division of labor: after agreeing on data dimensions to be tracked, each member created his or her own versions and iterations of the visualization. Then, students mutually showcased their progress and provided feedback, but then continued to work on own iterations. Only a few days before the deadline, contradicting or disliked approaches were completely discarded, and the group effort was bundled into towards one or two promising visualization candidates.

One straightforward limitation of that approach is certainly the amount of work done in vain, as students had to drop a large fraction of late-stage prototypes. On the other hand, such an approach allows them to explore a significantly larger design space due to the late convergence moment. Furthermore, students get used to letting their product go in favor of a more compelling team result.

\subsection{Limitations of the Experiment}
Although we were able to gather valuable data from the given responses, we recommend to consider the current findings as an intermediate result due to the low amount of pilot participants. Aspects such as memorability also need repeated evaluations in the future to measure the long-term effect strength. Furthermore, we suggest a more refined assessment of the overall course effectiveness, i.e., to measure how well the participants' visual literacy and visualization skills improved during the course.

\section{Conclusion and Future Work}

Our teaching experiment focused on introducing basic design thinking principles such as divergent thinking and brainstorming into graphics education by taking the \textit{Dear Data} method as a starting point. As we can see from the students' outcomes, the applied methodology encouraged participants to explore and learn a plethora of visualization techniques and proceed in a solution-oriented matter. The approach reduced the chance of falling into a tunnel-vision pattern that we often experienced in our previous lectures, i.e., the quick and sometimes misleading convergence toward one of the few well-known visualizations.

To assess the benefits and drawbacks of our teaching approach, we shared and discussed the outcomes of interviews and questionnaires from our pilot course. These results summarize our lessons learned regarding visualization novices, their decision making, and their creation process. We suggest to use these findings to further extend the computer graphics and visualization curriculum with the overarching goal to increase the overall visual literacy of our students. 

\section*{Acknowledgments}
This research was made possible in part by the NIH/ NINR - National Institute of Nursing Research Award Number R01NR014852.



\bibliographystyle{eg-alpha-doi}

\bibliography{ddata}


\end{document}